\documentclass[10pt,twocolumn,letterpaper]{article}

\usepackage{wacv}
\usepackage{times}
\usepackage{epsfig}
\usepackage{graphicx}
\usepackage{amsmath}
\usepackage{amssymb}
\usepackage{color}
\usepackage{subcaption}
\usepackage{animate}
\usepackage{balance}
\usepackage{float}
\usepackage{multirow}
\usepackage{animate}
\usepackage{algorithmicx}
\usepackage{algorithm}
\usepackage{algpseudocode}

\def\wacvPaperID{1012} 

\wacvfinalcopy 

\ifwacvfinal
\def\assignedStartPage{1} 
\fi

\usepackage[breaklinks=true,bookmarks=false]{hyperref}

\ifwacvfinal
\else
\fi

\begin{document}

\title{A Unified Framework for Compressive Video Recovery\\ from Coded Exposure Techniques}

\author{Prasan Shedligeri\textsuperscript{*}\\
Dept. of EE, IIT Madras\\
Chennai, India\\
{\tt\small ee16d409@ee.iitm.ac.in}
\and
Anupama S\textsuperscript{*}\\
Qualcomm India\\
Bangalore, India\\
\and
Kaushik Mitra \\
Dept. of EE, IIT Madras\\
Chennai, India\\
{\tt\small kmitra@ee.iitm.ac.in}
}

\maketitle
\begingroup\renewcommand\thefootnote{*}
\footnotetext{Equal contribution}
\endgroup

\begin{abstract}
Several coded exposure techniques have been proposed for acquiring high frame rate videos at low bandwidth.
Most recently, a Coded-2-Bucket camera has been proposed that can acquire two compressed measurements in a single exposure, unlike previously proposed coded exposure techniques, which can acquire only a single measurement.
Although two measurements are better than one for an effective video recovery, we are yet unaware of the clear advantage of two measurements, either quantitatively or qualitatively.
Here, we propose a unified learning-based framework to make such a qualitative and quantitative comparison between those which capture only a single coded image (Flutter Shutter, Pixel-wise coded exposure) and those that capture two measurements per exposure (C2B).
Our learning-based framework consists of a shift-variant convolutional layer followed by a fully convolutional deep neural network.
Our proposed unified framework achieves the state of the art reconstructions in all three sensing techniques.
Further analysis shows that when most scene points are static, the C2B sensor has a significant advantage over acquiring a single pixel-wise coded measurement. 
However, when most scene points undergo motion, the C2B sensor has only a marginal benefit over the single pixel-wise coded exposure measurement.
\end{abstract}

\vspace{-15pt}
\section{Introduction}

Cameras that can acquire high frame rate videos require high light sensitivity and massive data bandwidth increasing their cost significantly.
Hence, several methods have been proposed to first acquire a low frame rate video from a low-cost camera and computationally upsample the videos temporally  \cite{herbst2009occlusion,niklaus2017video,niklaus2017separable,jiang2018super}.
Computational imaging techniques have used compressive sensing theory to first acquire compressed video measurements at low bandwidth and then computationally reconstruct the high frame rate video signal \cite{baraniuk2017compressive}.
For visible light compressive video sensing, coded exposure techniques are the most popular ones with several compressive acquisition systems and reconstruction algorithms proposed over the years~\cite{raskar2006coded,gu2010coded,reddy2011p2c2,holloway2012flutter,llull2013coded,liu2013efficient,iliadis2018deep,yoshida2018joint,iliadis2020deepbinarymask,martel2020neural,li2020endtoend}.

\begin{figure}[t]
    \centering
    \setlength{\tabcolsep}{2pt}
    \begin{tabular}{cccc}
        \hline
         & Flutter & \multicolumn{1}{|c|}{Pixel-wise coded} & C2B (16x)\\
         & Shutter (8x)& \multicolumn{1}{|c|}{exposure (16x)} & (Two inputs)\\
        \hline
        \noalign{\smallskip}
        \rotatebox{90}{\hspace{20pt} Input}&
        \includegraphics[width=0.3\columnwidth]{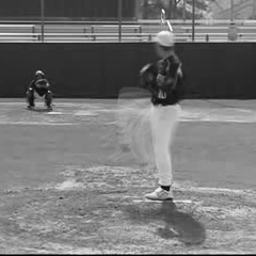}&
        \includegraphics[width=0.3\columnwidth]{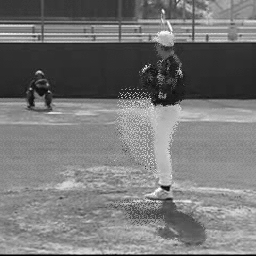}& 
        \includegraphics[width=0.3\columnwidth]{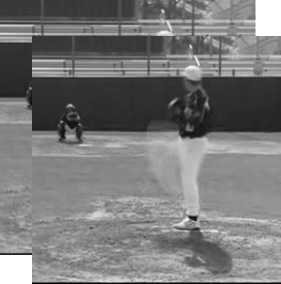}\\
        
        \rotatebox{90}{\hspace{2pt} Reconstruction} &
        \animategraphics[autoplay,loop,width=0.3\columnwidth]{5}{images/gmm-flutter8x/seq_03_recon_}{01}{08}&
        \animategraphics[autoplay,loop,width=0.3\columnwidth]{5}{images/ours-single/seq_03_recon_}{01}{16}&
        \animategraphics[autoplay,loop,width=0.3\columnwidth]{5}{images/ours-c2b/seq_03_recon_}{01}{16}\\
        
         & 27.82 dB, 0.908 & 32.29 dB, 0.946 & \bf 34.65 dB, 0.972\\
        
    \end{tabular}
    \vspace{-5pt}
    \caption{We propose a unified deep learning based framework that allows us to compare the performance of various coded exposure techniques. We show reconstruction results for a video sequence from \emph{DNN test set}~\cite{yoshida2018joint}. (videos can be viewed in Adobe Reader)}
    \label{fig:intro}
    \vspace{-20pt}
\end{figure}

In coded exposure techniques, a pre-determined code is used to multiplex the temporal dimension of the video signal into compressed measurements.
Recently, a novel prototype sensor based on multi-bucket pixels named Coded-2-Bucket sensor~\cite{sarhangnejad20195,wei2018coded} was introduced.
While allowing for per-pixel control of the ``shutter'', this sensor also can acquire two compressed measurements per exposure using the $2$ light-collecting buckets per pixel.
Based on the number of measurements acquired per exposure, we can classify these sensing techniques into two categories: a) single compressed measurement (such as flutter shutter and pixel-wise coding)~\cite{holloway2012flutter,raskar2006coded,llull2013coded,reddy2011p2c2,liu2013efficient,iliadis2018deep,yoshida2018joint,iliadis2020deepbinarymask,martel2020neural,li2020endtoend} and b) two compressed measurements per exposure~\cite{sarhangnejad20195}.
It is expected that two measurements should lead to better video reconstruction quality compared to a single measurement.
However, the performance improvement provided by two compressed measurements over a single compressed measurement is yet to be investigated.
As the C2B sensor is recently introduced, no previous algorithm exists, making a quantitative comparison between the single and two compressed measurement techniques.
While \cite{li2020endtoend} uses C2B, only a qualitative comparison on a single video sequence is made for single and two measurement cases.
An extensive quantitative and qualitative comparison has not been made, and it can help determine how much advantage is gained by acquiring two measurements over just one.
This comparison of the different sensing architectures will also provide users with a tool to determine which sensing technique is better for a given scenario.

With this objective, we propose a unified learning-based framework with which we wish to make an extensive evaluation.
This learning based framework should be usable for recovering videos from various single and two measurement techniques, particularly, Flutter-Shutter \cite{raskar2006coded}, pixel-wise coded exposure \cite{veeraraghavan2010coded,gupta2010flexible} and C2B \cite{sarhangnejad20195}.
Most of the previously proposed algorithms for compressive video recovery use fully connected networks, and, ideally, we can use any of those networks for our framework.
However, fully connected networks have fallen out of favour for most image processing tasks as they have a large number of trainable parameters and are also hard to scale up for large spatial/temporal resolutions.
Hence, we design our framework to be fully-convolutional, enabling reconstruction of the full resolution video sequence in a single forward pass.
In \cite{martel2020neural} it has also been demonstrated that a fully convolutional network provides better reconstruction results than fully connected networks.
Later, we provide an intuitive explanation for why a convolutional network with local spatial connectivity is actually more suitable for this problem than fully connected networks with global connectivity.
Our framework also uses the recently proposed \emph{shift-variant} convolutional (SVC) layer \cite{okawara2020action} that has shown to be effective for feature extraction from a coded image input.

The proposed algorithm is divided into two stages, where the first stage uses the SVC layer for an exposure code aware feature extraction.
In the second stage, a deep, fully convolutional neural network is used to learn the non-linear mapping to the full resolution video sequence.
Extensive comparisons show that our proposed learning framework provides state of the art results on all three sensing techniques.
Using our unified framework, we quantitatively evaluate the performance of the various coded exposure techniques.
As expected, pixel-wise coded exposure techniques produce much better video reconstructions than global coded exposure technique such as FS. 
We also confirm that acquiring two compressed measurements as in C2B is better than capturing just a single compressed measurement.
The advantage is significant for a largely stationary scene (Fig.~\ref{fig:c2bvsingle}).
However, C2B is only marginally beneficial over a single pixel-wise coded compressed measurement when most scene points undergo motion.

In summary we make the following contributions:
\begin{itemize}
    \item[\textbullet] We provide a deep learning framework using which various coded exposure techniques can be compared.
    \vspace{-7pt}
    \item[\textbullet] We make an extensive quantitative and qualitative evaluation of the different coded exposure techniques for compressive video recovery.
    \vspace{-7pt}
    \item[\textbullet] Our proposed approach matches or exceeds the state-of-the-art video reconstruction algorithms for each of the sensing techniques.
    \vspace{-7pt}
    \item[\textbullet] We show that C2B has significant advantage over per-pixel exposure coding in reconstructing videos of scenes consisting of significant static regions.
\end{itemize}

\section{Related Work}

\textbf{High speed imaging techniques with conventional sensor:}
Conventional image sensors capture a sharp video by using exposures shorter than the sampling period of the video.
Frame interpolation techniques~\cite{herbst2009occlusion,niklaus2017video,jiang2018super,niklaus2017separable} can be used to interpolate multiple frames between any two acquired frames and thereby increasing the video frame rate.
When a long exposure is used, a blurred frame is acquired which encodes the full motion information.
Recent learning based methods~\cite{purohit2019bringing,jin2018learning} have been used to decode the motion information from a single blurred frame into multiple video frames.

\noindent\textbf{Computational Imaging techniques:}
For scenes with little to no depth variations techniques using arrays of low-cost low frame rate cameras have shown to be effective at computationally recovering the high frame rate video~\cite{wilburn2005high,shechtman2005space,agrawal2010optimal}.
A hybrid imaging system which uses one low-frame rate but high spatial resolution and another high frame rate but low spatial resolution sensors has been proposed for image deblurring~\cite{nayar2004motion} and high spatio-temporal resolution video recovery~\cite{paliwal2020deep}.
Recently, a hybrid imaging system consisting of image and event sensor has been proposed for high speed image reconstruction~\cite{shedligeri2019photorealistic, wang2019event, wang2020joint}.

Motivated from the compressive sensing theory, several imaging architectures have been proposed for video compressive sensing problem~\cite{baraniuk2017compressive}.
Flutter shutter is a global exposure coding technique which was first introduced for motion deblurring~\cite{raskar2006coded} and then extended for video recovery from the compressed measurements~\cite{holloway2012flutter}.
A pixel-wise coded exposure system was proposed in \cite{reddy2011p2c2} which demonstrated the recovery of high temporal resolution video from measurements compressed using spatial light modulator.
A per-pixel control of the exposure was shown in \cite{liu2013efficient}, using only a commercially available CMOS image sensor without the need for any other hardware.
The recently introduced multi-bucket sensors such as \emph{Coded-2-Bucket} cameras~\cite{sarhangnejad20195,wei2018coded}, have reduced the complexity of per-pixel exposure control to a great extent.
As video recovery from the compressed measurements is an ill-posed problem, strong signal priors are necessary for solving the inverse problem.
While analytical priors such as wavelet domain sparsity~\cite{reddy2011p2c2,park2009multiscale}, TV-regularization~\cite{yuan2016generalized} have been used, learning based algorithms such as Gaussian mixture models~\cite{yang2014video}, dictionary learning~\cite{liu2013efficient} and neural network based models~\cite{iliadis2018deep,iliadis2020deepbinarymask,yoshida2018joint} have shown better performance than analytical priors.
While many of the deep learning based methods use fully connected networks for the signal recovery, a very recent paper~\cite{li2020endtoend} uses a fully convolutional network to learn a denoising prior to iteratively solve the inverse problem.

\begin{figure*}[ht]
    \centering
    \includegraphics[width=0.8\textwidth]{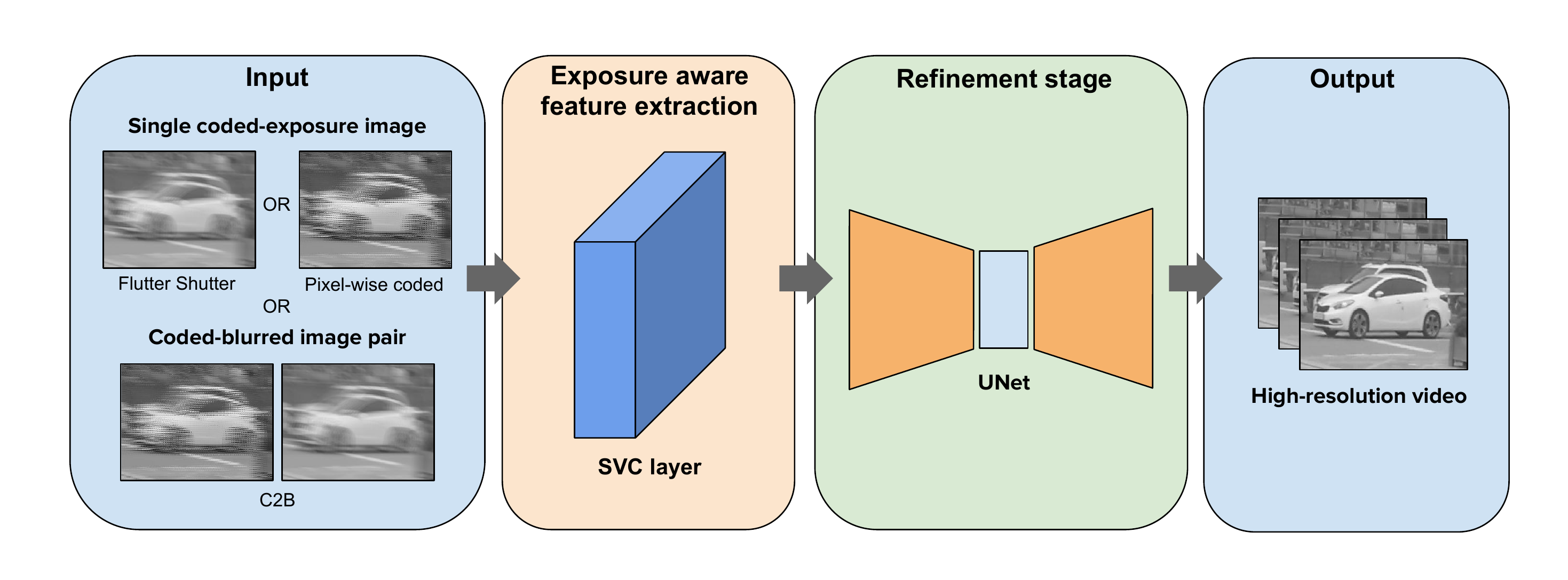}
    \vspace{-5pt}
    \caption{Our proposed algorithm takes in compressed measurements from the different coded exposure techniques as input and output full spatial and temporal resolution video in a single forward pass. Our proposed algorithm is fully convolutional and consists of a feature extraction stage and a refinement stage.}
    \label{fig:arch}
    \vspace{-15pt}
\end{figure*}

\section{A Unified Framework for Compressive Video Recovery Using Fully Convolutional Network}
\label{sec:method}
In this section, we elaborate on our proposed method to obtain the video signal from its compressed measurements.
Our proposed algorithm takes in as input the compressed video measurements and outputs the video sequence at full spatial and temporal resolution in a single forward pass.
The proposed architecture consists of two stages, as shown in Fig.~\ref{fig:arch}.
First, features are extracted from the compressed measurements using an exposure aware feature extraction stage consisting of shift variant convolutional layer.
In the second stage, a deep neural network takes in the extracted features and outputs the full resolution video sequence.
Our network architecture is flexible enough that it can be used for video reconstruction from all three coded exposure techniques considered here.
All we need to do is train the network for these different inputs.

In Sec.~\ref{sec:sensing}, we provide motivation for using CNN for extracting relevant features from the compressed measurements.  
In Sec.~\ref{sec:inversion}, we elaborate on the use of \emph{shift-variant} convolutional layer for handling pixel-wise coded exposure measurements and in Sec.~\ref{sec:refine} we specify the loss function used in the training our network.

\subsection{Motivation for Using CNN}
\label{sec:sensing}
Several previous learning-based algorithms for compressive video recovery from coded exposure techniques have used fully connected networks~\cite{yoshida2018joint,iliadis2020deepbinarymask}.
In \cite{martel2020neural}, it has been shown that a fully convolutional network provides better reconstruction than fully connected networks for compressive video sensing.
This section shows that a fully convolutional network is a better choice for solving our problem than a fully connected network.

For coded exposure techniques, each pixel in the compressed measurement is a linear combination of the underlying video sequence at that pixel alone.
As there is no spatial multiplexing involved, it is possible to recover the video sequence at each pixel independently of the neighboring pixels.
However, by using the information in a small neighborhood of a pixel, we can exploit the spatio-temporal redundancy inherent in natural video signals.
Fully connected networks that are used in previous works provide global connectivity at the cost of much larger computational complexity and learning parameters. 
Thus, they should be used for solving inverse problems where global multiplexing occurs in the forward model, such as FlatCam~\cite{asif2016flatcam}.
With a toy example and elementary mathematical operations, we demonstrate next that fully connected networks with global connectivity are overkill and fully convolutional network with local spatial connectivity is a better design choice for our problem.

\subsubsection{Toy example demonstration}
Consider a video signal $x$ of size $H\times W\times T$ with $x_t$ representing each of the $T$ frames of the video signal.
A binary exposure sequence $\phi$ of dimension $H\times W\times T$ is used for temporally multiplexing the signal $x$ into the measurement $I$.
Mathematically, we can write the forward model as:
\begin{align}
    \label{eq:sum_of_exposures}
    I = \sum_{t=1}^T \phi_t\odot x_t ~,
\end{align}
where $\phi_t$ represents the code corresponding to each frame of $\phi$ and $\odot$ represents element-wise multiplication.

The linear system in Eq.~\eqref{eq:sum_of_exposures} can be represented in the matrix-vector form as follows:
\begin{align}
    I = \Phi X ~,
\end{align}
where $\Phi$ is a matrix representation of $\phi$ and $X$ is a column vector obtained by vectorizing $x$.
The minimum $L_2$-norm solution for the signal $X$ can be obtained by:
\begin{align}
    \label{eq:pinv_problem}
    \min_X \|X\|_2 \\
    s.t. ~ I = \Phi X ~ .
\end{align}
Note that there are better reconstruction techniques such as dictionary learning which uses $L_0$ or $L_1$ norm on sparse transform coefficients of $X$~\cite{liu2013efficient}.
But here our main goal is to show that CNN is appropriate for solving our inverse problem and hence we only provide a justification with $L_2$-norm, that has a closed-form solution.
The approximate solution $\tilde X$ for Eq.~\eqref{eq:pinv_problem} is given by,
\begin{align}
    \label{eq:pinv_soln}
    \tilde X &= \Phi^{\dagger} I ~ , \\
    \Phi^{\dagger} &= \Phi^T(\Phi\Phi^T)^{-1} .
\end{align}

\begin{figure*}
    \centering
    \includegraphics[width=0.8\textwidth]{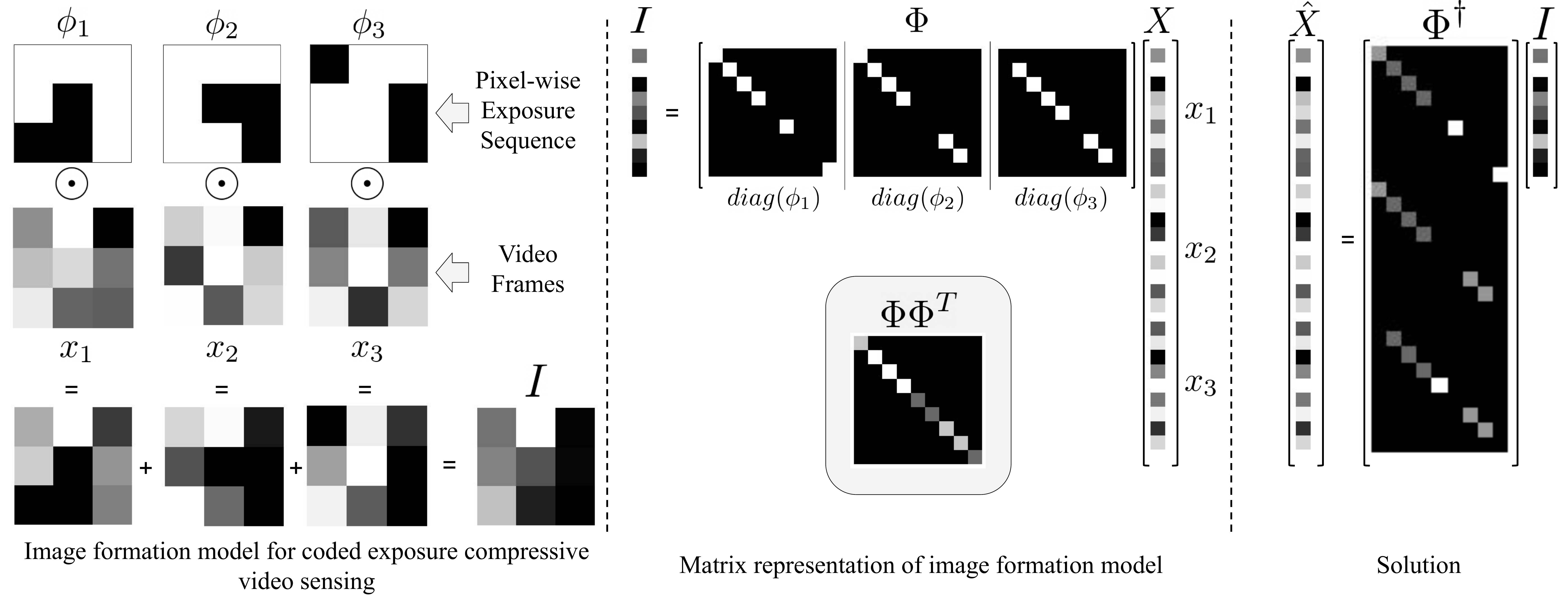}
    \caption{We show a toy example of pixel-wise coded exposure technique for compressing a video sequence of size $3\times 3\times 3$. $\Phi$ and $X$ are the matrix and vector representation of the exposure sequence $\phi$ and the video sequence $x$, respectively. From the pseudo-inverse solution we see that the temporal video sequence reconstruction at any pixel depends only on the measurement and the code at that pixel alone. This motivates our choice of a fully convolutional design.
    }
    \label{fig:model}
    \vspace{-10pt}
\end{figure*}

We notice that the matrix $\Phi\Phi^T$ is a diagonal matrix of dimension $HW\times HW$, and so is the matrix $(\Phi\Phi^T)^{-1}$.
As shown in Fig.~\ref{fig:model}, the matrix $\Phi^{\dagger}$ is the matrix $\Phi^T$ whose columns are scaled by the entries of the diagonal matrix $(\Phi\Phi^T)^{-1}$.
From the solution shown in Fig~\ref{fig:model}, it is clear that the temporal sequence at each pixel of the video is recovered only from the compressed measurement captured at that pixel. 
For example, if we consider the $j^{th}$ pixel location, then the estimated temporal sequence $(\hat x^j_1,\hat x^j_2,\hat x^j_3)$ corresponding to the $j^{th}$ pixel location depends only on the compressed measurement at the same pixel location $I^j$. 
This shows that CNN with local connectivity is more than sufficient for video reconstruction from coded exposure images.

\subsection{Feature Extraction Using Shift-Variant Convolution (SVC)}
\label{sec:inversion}
In Sec.~\ref{sec:sensing}, we determined that to recover a video at a particular pixel, only that pixel's compressed measurements are necessary.
Hence, the local connectivity offered by CNNs can be efficiently used for the task of recovering the underlying video signal.
However, CNNs share the same weights across the whole input image. 
In pixel-wise coded exposure, the compressed measurement can be encoded using a different exposure sequence at each pixel.
From Eq.~\eqref{eq:pinv_soln} and Fig.~\ref{fig:model}, we see that the estimated video sequence at a particular pixel is dependent on the exposure sequence at that particular pixel.
Hence, for pixels with different exposure sequence, using a different set of weights in the convolutional layer is desirable.

In flutter shutter video camera, each pixel in the image shares the same coded exposure sequence. Hence, the same learned convolutional weights $w$ can be used to recover the underlying video signal for all the pixels.
Thus, for recovering video sequences from the flutter shutter camera, we build our inversion stage as a standard convolutional layer as it achieves the functions mentioned above: local connectivity and shared weights across the whole image.

In pixel-wise coded exposure and C2B architectures, the underlying coded exposure sequence can change from one pixel to the next. 
In practice, a predetermined code of size $m\times n \times T$ is repeated over the entire image with a stride of $m\times n$ pixels.
Hence, a standard convolutional layer cannot be directly used as it shares the same set of weights across the whole image.
Instead, a convolutional layer, which can share weights for every $m\times n^{th}$ pixel, is desirable.
Such a convolutional layer whose weights \emph{vary} in a local neighborhood of $m\times n$ pixels was proposed in \cite{okawara2020action} called \emph{shift-variant} convolutional (SVC) layer.
This layer allows the network the freedom to learn different weights to invert the linear system when the underlying exposure sequence is different.
Hence, we use this layer to extract adaptive features from the input compressed measurement. 
These extracted features are input to the next stage of the network, which predicts the full resolution video sequence.











\subsection{Refinement Stage}
\label{sec:refine}

The refinement stage takes as input the features extracted from the \emph{shift-variant} convolutional layer and outputs a refined video sequence $\hat X$.
Our refinement stage consists of a UNet~\cite{ronneberger2015u} like deep neural network. 
Our proposed Unet model consists of $3$ encoder stages followed by a bottleneck layer and $3$ decoder stages. In each of the encoder stages, the feature maps are downsampled spatially by a factor of $2$ and upsampled by the same factor in corresponding decoder stage.
The output of this network is supervised using $L_1$ loss function. We also add a TV-smoothness loss on the final predicted video sequence.
Our overall loss function then becomes,
\begin{equation}
    \begin{aligned}
    \mathcal{L} &= \mathcal{L}_{ref} + \lambda_{tv} \mathcal{L}_{tv} \\
    \mathcal{L}_{ref} &= \| \hat X - X \|_1  \\
    \mathcal{L}_{tv} &= \| \nabla \hat X  \|_1
\end{aligned}
\end{equation}
where $\nabla$ is the gradient operator in the x-y directions and $\lambda_{tv}$ weights the smoothness term in the overall loss function.

\begin{figure}\footnotesize
    \centering 
    \setlength{\tabcolsep}{2pt}
    \begin{tabular}{ccc}
        \hline
        \multicolumn{3}{c}{Flutter shutter (8x)}\\
        \hline
        Input& GMM~\cite{yang2014video}& Ours\\
        \includegraphics[width=0.2\columnwidth]{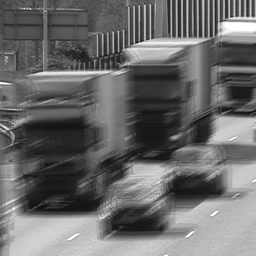}&
        \animategraphics[autoplay,loop,width=0.2\columnwidth]{5}{images/gmm-flutter8x/pred_11_frame_}{01}{08}& 
        \animategraphics[autoplay,loop,width=0.2\columnwidth]{5}{images/ours-flutter8x/seq_11_recon_}{01}{08}\\
        & 17.46, 0.586& \bf 21.82, 0.773\\
    \end{tabular}

    \begin{tabular}{ccccc}
        \hline
        \multicolumn{5}{c}{Pixel-wise coded exposure (16x)}\\
        \hline
        \noalign{\smallskip}
        Input & GMM~\cite{yang2014video} & DNN~\cite{yoshida2018joint} & AAUN\cite{li2020endtoend} & Ours\\

        \includegraphics[width=0.18\columnwidth]{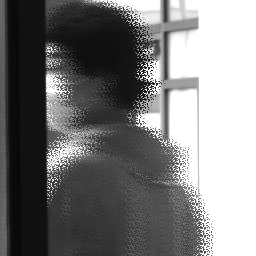} &
        \animategraphics[autoplay,loop,width=0.18\columnwidth]{5}{images/gmm/pred_14_frame_}{01}{16} &
        \animategraphics[autoplay,loop,width=0.18\columnwidth]{5}{images/eccv/wa_recon}{01}{16} &
        \animategraphics[autoplay,loop,width=0.18\columnwidth]{5}{images/iccp/iccp_014_}{00}{15}&
        \animategraphics[autoplay,loop,width=0.18\columnwidth]{5}{images/ours-single/seq_14_recon_}{01}{16} \\
        
        & 31.54, 0.937 & 31.88, 0.94 & 32.99, 0.960 & \bf 34.03, 0.963\\

        \includegraphics[width=0.18\columnwidth]{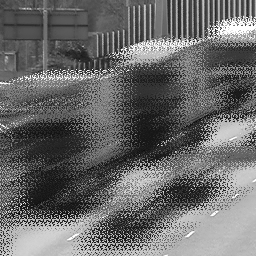} & \animategraphics[autoplay,loop,width=0.18\columnwidth]{5}{images/gmm/pred_11_frame_}{01}{16} &
        \animategraphics[autoplay,loop,width=0.18\columnwidth]{5}{images/eccv/Traffic_recon}{01}{16} &
        \animategraphics[autoplay,loop,width=0.18\columnwidth]{5}{images/iccp/ours_011_}{00}{15} &
        \animategraphics[autoplay,loop,width=0.18\columnwidth]{5}{images/ours-single/seq_11_recon_}{01}{16}\\
        & 22.25, 0.747 & 22.69, 0.764 & 23.75, 0.8 & \bf 24.20, 0.828\\
        \noalign{\smallskip}
        \end{tabular}

        \begin{tabular}{cccc}
        \hline
        \multicolumn{4}{c}{Coded-2-bucket exposure (16x)}\\
        \hline
        \noalign{\smallskip}
        Coded image & Blurred image & GMM~\cite{yang2014video} & Ours\\
        \includegraphics[width=0.18\columnwidth]{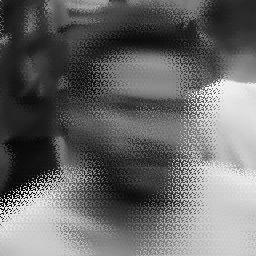}& 
        \includegraphics[width=0.18\columnwidth]{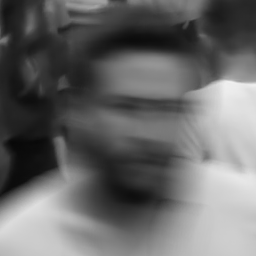}& 
        \animategraphics[autoplay,loop,width=0.18\columnwidth]{5}{images/gmm-c2b-gopro/pred_03_frame_}{01}{16}& \animategraphics[autoplay,loop,width=0.18\columnwidth]{5}{images/ours-c2b-gopro/seq_03_recon_}{01}{16}\\
        & & 33.51, 0.959 & \bf 35.35, 0.972\\
        \includegraphics[width=0.18\columnwidth]{images/input/seq_11_coded.png}& \includegraphics[width=0.18\columnwidth]{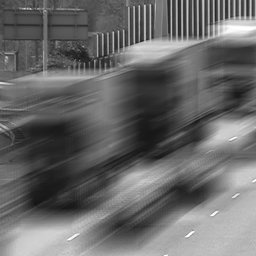}& \animategraphics[autoplay,loop,width=0.18\columnwidth]{5}{images/gmm-c2b/pred_11_frame_}{01}{16}& \animategraphics[autoplay,loop,width=0.18\columnwidth]{5}{images/ours-c2b/seq_11_recon_}{01}{16}\\
        & & 23.17, 0.779 & \bf 24.93, 0.851\\
    \end{tabular}
    \caption{Visual comparison of reconstructed videos for different coded exposure techniques and reconstruction algorithms. Our proposed method performs better than the existing methods GMM~\cite{yang2014video}, DNN~\cite{yoshida2018joint} AAUN~\cite{li2020endtoend}, and also doesn't suffer from block artifacts caused by patch-wise reconstruction. As expected, C2B produces better results than pixel-wise coded imaging. FS lags far behind.
    (videos can be viewed in Adobe Reader)}
    \label{fig:analysis}
    \vspace{-15pt}
\end{figure}

\section{Experimental Results}

\subsection{Experimental and Training Setup}
\label{sec:exptsetup}
\textbf{Ground truth data preparation:} 
We trained our proposed network using GoPro dataset~\cite{Nah_2017_CVPR} consisting of $22$ video sequences at a frame rate of $240$ fps and spatial resolution of $720\times 1280$.
The first $512$ frames from each of the $22$ sequences are spatially downsampled by $2$ for preparing the training data.
Overlapping video patches of size $64\times 64\times 16$ (height$\times$width$\times$frames) are extracted from the video sequences by using a sliding 3D window of $(32,32,8)$ pixels resulting in $263,340$ training patches.
Similarly, for 8-frame reconstruction, we extracted video patches of size $64\times 64\times 8$ and shifting the window by $(32,32,4)$ pixels.
The network was trained in PyTorch~\cite{paszke2019pytorch} using Adam optimizer~\cite{kingma2014adam} with a learning rate of $0.0001$, $\lambda_{tv}$ of $0.1$ and batch size of $50$ for $500$ epochs\footnote{ \url{https://github.com/asprasan/unified\_framework}}.

\noindent\textbf{Network architecture for each sensing technique:} 
We trained our network separately for each of the different coded exposure techniques - \emph{Flutter Shutter (FS), Pixel-wise coded exposure}, and \emph{Coded-2-Bucket}. 
For FS, we trained our proposed network for 16-frame reconstruction and 8-frame reconstruction.
As FS uses global code, a standard convolutional layer is used as a feature extraction layer in place of the SVC layer.
We use the SVC layer as described in Sec.~\ref{sec:inversion} as a feature extraction stage for pixel-wise coded exposure and C2B.

\noindent\textbf{Input to the network:}
In the case of FS, the input to the network is a single coded exposure image obtained by multiplexing with a global exposure code.
We used the exposure code obtained by maximizing the minimum of the DFT values' magnitude and minimizing the variance of the DFT values~\cite{raskar2006coded}, over all possible binary codes.
For the case of pixel-wise coded exposure, the coded mask of size $8\times 8\times 16$ is repeated spatially to make it the same dimension as input, which is then used for multiplexing.
We used the \emph{optimized SBE mask} exposure code proposed in \cite{yoshida2018joint} for this purpose.
In the case of C2B exposure, the input to the network can either be a pair of coded and complement-coded images or a pair of coded and fully-exposed images.
The output of the C2B sensor is two images that are coded using complementary exposure sequences (i.e., $\phi$ and $1 - \phi$).
We used the same exposure pattern \emph{optimized SBE mask} from \cite{yoshida2018joint} for C2B exposure as well.
The fully-exposed or blurred image is obtained by adding the coded and complementary coded images.
The image pair for the C2B sensor are stacked as two channels and provided as input to the proposed algorithm.

\setlength{\tabcolsep}{3pt}
\begin{table}
    \fontsize{9}{7.2}\selectfont
    \begin{center}
    \begin{tabular}{llll}
        \hline
        \noalign{\smallskip}
        Exposure\qquad & Algorithm & \multicolumn{2}{c}{Test data}\qquad\\

        \noalign{\smallskip}
        \hline
        \noalign{\smallskip}

        & & DNN set~\cite{yoshida2018joint}\qquad& GoPro set~\cite{Nah_2017_CVPR} \\
        \cline{3-4}
        \noalign{\smallskip}

        \multirow{2}{*}{FS 8x} & GMM~\cite{yang2014video} & 23.90, 0.818& 23.30, 0.766 \\

        & Ours & \bf 24.06, 0.833 & \bf 25.03, 0.811\\

        \noalign{\medskip}
        
        \multirow{2}{*}{FS 16x} & GMM~\cite{yang2014video} & 21.50, 0.738 &  21.45, 0.697 \\

        & Ours & \bf 21.69, 0.752 & \bf 21.61, 0.710 \\
        \noalign{\smallskip}

        \hline
        \noalign{\smallskip}
        & & DNN set~\cite{yoshida2018joint}\qquad& GoPro set~\cite{Nah_2017_CVPR} \\
        \cline{3-4}
        \noalign{\smallskip}
        & GMM~\cite{yang2014video}& 29.31, 0.898 & 29.94, 0.887 \\
        Pixel-wise & DNN~\cite{yoshida2018joint}& 30.21, 0.905 & 30.27, 0.890 \\
        coded 16x & AAUN~\cite{li2020endtoend} & 28.5, 0.882 & 31.6, 0.910 \\
        & Ours & \bf 31.14, 0.925 & \bf 31.76, 0.914 \\

        \noalign{\smallskip}
        \hline
        \noalign{\smallskip}
        & & DNN set~\cite{yoshida2018joint}\qquad& GoPro set~\cite{Nah_2017_CVPR} \\
        \cline{3-4}
        \noalign{\smallskip}
        \multirow{2}{*}{C2B 16x}& GMM~\cite{yang2014video}& 30.94, 0.914 & 30.84, 0.898 \\
        & Ours & \bf 32.23, 0.935 & \bf 32.34, 0.920 \\
        \noalign{\smallskip}
        \hline
    \end{tabular}
    \end{center}
    \caption{Quantitative results for different coded exposure techniques and reconstruction algorithms. The table lists average PSNR(dB) and SSIM of reconstructed videos from \emph{DNN set}~\cite{yoshida2018joint} and \emph{GoPro set}~\cite{Nah_2017_CVPR}.{\smallskip}}
    \label{table:analysis}
    \vspace{-5pt}
\end{table}
\setlength{\tabcolsep}{1.4pt}

\setlength{\tabcolsep}{2pt}
\begin{table}
    \centering
    \begin{tabular}{ccccc}
        \hline
        Exposure & \multicolumn{4}{c}{CPU run-time (GPU run-time) in seconds}\\
        \hline 
         & GMM~\cite{yang2014video} & DNN~\cite{yoshida2018joint} & AAUN~\cite{li2020endtoend} & Ours\\
        \cline{2-5}
        Pixel-wise & 78.7 & 4.6 (2.7) & 11.1 (0.3) & 3.6 (0.011)\\
        C2B & 96.4 & -- & -- & 4.1 (0.013)\\
    \end{tabular}
    \caption{Run time for various algorithms to reconstruct a single $256\times 256\times 16$ frame sequence. For algorithms that are accelerated by GPU, the run times are provided in parentheses. The run times are for an Intel i7 CPU and Nvidia GeForce 2080 Ti GPU.}
    \label{tab:my_label}
    \vspace{-12pt}
\end{table}

\subsection{Analysis of Video Reconstruction for Various Compressive Sensing Systems}
\label{sec:analysis}

In this section, we qualitatively and quantitatively assess video reconstruction from compressed measurements captured by different coded exposure techniques - \emph{FS}, \emph{pixel-wise coded exposure}, and \emph{C2B}.  
We compared our proposed method with existing state-of-the-art algorithms for video reconstruction such as GMM-based inversion \cite{yang2014video}, DNN \cite{yoshida2018joint} and AAUN \cite{li2020endtoend}. 
We used two sets of test videos with a different spatial resolution to perform this analysis. First, we used the test set that was used for evaluation in \emph{DNN~\cite{yoshida2018joint}}, consisting of 14 videos of spatial resolution $256\times 256$ and $16$ frames each. 
For the second set, we randomly selected 15 videos of resolution $720\times 1280$ and $16$ frames each, from the \emph{GoPro test dataset~\cite{Nah_2017_CVPR}}.

For FS, we compared our proposed method with the GMM-based video reconstruction method~\cite{yang2014video} for 8-frame and 16-frame reconstructions.
For single pixel-wise coded exposure sensing, we compare with GMM-based inversion~\cite{yang2014video} and state-of-the-art deep learning based methods, DNN~\cite{yoshida2018joint} and AAUN~\cite{li2020endtoend}, for 16-frame reconstruction. 
For C2B exposure, we compare with GMM-based inversion~\cite{yang2014video} for 16-frame reconstruction from a pair of coded and blurred images. 
We trained the GMM~\cite{yang2014video} model with $20$ components using the same training dataset as described in Sec.~\ref{sec:exptsetup}. 
We used $8\times 8\times 8$ patches to train the GMM~\cite{yang2014video} for 8-frame reconstruction and $8\times 8\times 16$ patches for 16-frame reconstruction.
We used the pre-trained model for DNN proposed in \cite{yoshida2018joint}. 
We trained the AAUN~\cite{li2020endtoend} algorithm on the same training dataset as described in Sec.~\ref{sec:exptsetup}. The model was trained for $80$ epochs on patches of size $128\times 128$ for $16$-frame reconstruction.

\textbf{Comparison analysis:} Qualitative reconstruction results are shown in Fig.~\ref{fig:analysis} and quantitative results are summarized in Table~\ref{table:analysis}.
FS produces satisfactory results for $8$-frame reconstruction but struggles to reconstruct $16$ frames.
Pixel-wise coded exposure can perform $16$-frame reconstruction with good fidelity.
For natural images, the intensities in a small spatial neighborhood are correlated.
Intuitively, using different exposure sequences for different pixels, is equivalent to making multiple measurements, which helps in recovering the information better.
As our algorithm exploits the spatial correlation structure, the pixel-wise coded exposure technique will have an advantage over the global, flutter shutter imaging technique in the fidelity of the reconstructed video.
The C2B exposure provides an additional advantage by capturing information that is lost by the pixel-wise coded exposure and hence produces better reconstruction than pixel-wise coded exposure.
Overall, we observe a similar trend in the reconstruction performance of different sensing techniques in both GMM~\cite{yang2014video} our proposed and model.
We see that, overall, C2B provides the best reconstruction and FS performs the worst, while there is only a slight quantitative advantage for C2B when compared to pixel-wise exposure.
We further compare the performance of pixel-wise coded exposure with C2B exposure in the following section.

Our proposed fully-convolutional model performs better than the existing methods, GMM~\cite{yang2014video}, DNN~\cite{yoshida2018joint} and AAUN~\cite{li2020endtoend}, for all the sensing techniques.
Since we reconstruct the full video, our proposed method doesn't suffer from block artifacts, which is seen in patch-wise reconstruction methods such as GMM and DNN.
A comparison of run times of various algorithms on CPU as well as GPU has also been provided in Table~\ref{tab:my_label}.
Patch-based reconstruction methods such as GMM and DNN require a significantly longer time to reconstruct a single video sequence compared to AAUN~\cite{li2020endtoend} and our algorithm.
Being an iterative deep learning algorithm, AAUN~\cite{li2020endtoend} takes 3x and 10x longer time than our proposed algorithm on CPU and GPU, respectively.

\setlength{\tabcolsep}{1pt}
\begin{figure}
    \centering
    \begin{tabular}{cc|cc}
        \hline
        single pixel-wise & \multirow{2}{*}{C2B} & single pixel-wise & \multirow{2}{*}{C2B}\\
        coded exposure &  & coded exposure & \\
        \hline
        \noalign{\smallskip}
        \multicolumn{2}{c|}{purely dynamic scene}&
        \multicolumn{2}{c}{partly dynamic scene}\\
        \noalign{\smallskip}
        \animategraphics[autoplay,loop,width=0.2\columnwidth]{5}{images/ours-single/seq_05_recon_}{01}{16}&  
        \animategraphics[autoplay,loop,width=0.2\columnwidth]{5}{images/ours-c2b/seq_05_recon_}{01}{16}& 
        \animategraphics[autoplay,loop,width=0.2\columnwidth]{5}{images/ours-single/seq_13_recon_}{01}{16}&
        \animategraphics[autoplay,loop,width=0.2\columnwidth]{5}{images/ours-c2b/seq_13_recon_}{01}{16}\\
        29.95, 0.904 & \bf 30.38, 0.908 & 32.21, 0.954 & \bf 34.50, 0.970\\ 
        \hline
        \multicolumn{2}{c|}{largely stationary scene}& \multicolumn{2}{c}{largely stationary scene}\\
        \animategraphics[autoplay,loop,width=0.2\columnwidth]{5}{images/pets-single/seq_07_recon_}{01}{16}& \animategraphics[autoplay,loop,width=0.2\columnwidth]{5}{images/pets-c2b/seq_07_recon_}{01}{16}& \animategraphics[autoplay,loop,width=0.2\columnwidth]{5}{images/pets-single/seq_06_recon_}{01}{16}&         \animategraphics[autoplay,loop,width=0.2\columnwidth]{5}{images/pets-c2b/seq_06_recon_}{01}{16}\\
        27.53, 0.914 & \bf 33.07, 0.977 & 28.11, 0.917 & \bf 35.48, 0.980\\
        \hline
    \end{tabular}
    \caption{Videos reconstructed using our proposed method. (a) and (b) are from \textit{DNN set~\cite{yoshida2018joint}}, (c) and (d) are cropped videos from \textit{PETS2009 dataset~\cite{ferryman2009pets2009}}. For videos containing significant stationary regions, C2B produces much better reconstructions than single coded exposure image, but the improvement is not significant for purely dynamic videos. (videos can be viewed in Adobe Reader)}
    \label{fig:c2bvsingle}
    \vspace{-15pt}
\end{figure}

\subsection{When Does C2B Have a Significant Advantage over Pixel-wise Coded Exposure?}
\label{sec:c2bvsingle}
In Sec.~\ref{sec:analysis}, we observe that C2B based sensing provides only a slight advantage compared to pixel-wise coded exposure technique.
To analyze and identify the cases where C2B provides a significant advantage over pixel-wise coded exposure, we conduct experiments on different kinds of videos: purely dynamic sequences, partly-dynamic-partly-static sequences, and largely static sequences.
We use our proposed method to compare video reconstruction from a pixel-wise coded exposure image and from a coded-blurred image pair obtained from C2B.
We explain why we use a blurred image with the coded image as input through an ablation study in Sec~\ref{sec:ablation}.
Fig.~\ref{fig:c2bvsingle} shows reconstructed results for the different cases of video sequences mentioned above.
For purely dynamic scenes, C2B does not show a notable performance improvement over pixel-wise coded exposure. 
However, for videos containing significant static regions, C2B produces much better reconstruction results than pixel-wise coded exposure. 
If we consider a scene composed of both stationary and dynamic regions, the dynamic regions are better captured by the coded exposure image, while the stationary regions are better captured by the fully-exposed image.
Therefore, it follows that videos containing stationary regions can be better recovered by using the additional information captured by C2B.

\subsection{Ablation Study on Proposed Architecture and C2B Input}
\label{sec:ablation}

\begin{figure}[t]
    \begin{center}
    \begin{tabular}{c|cc|c}
        \hline
        \multirow{2}{*}{Input} & \multicolumn{2}{|c|}{SVC(16)+U-Net} & SVC(64)+U-Net\\
        \cline{2-4}
        & \multicolumn{1}{|c}{Intermediate} & \multicolumn{1}{c|}{Final} & Final\\
        \includegraphics[width=0.2\columnwidth]{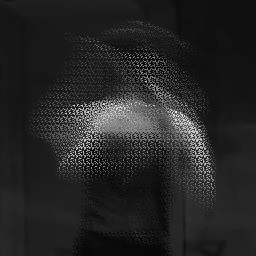}& \animategraphics[autoplay,loop,width=0.2\columnwidth]{5}{images/ours-single-interm/seq_08_interm_}{01}{16} & \animategraphics[autoplay,loop,width=0.2\columnwidth]{5}{images/ours-single-interm/seq_08_recon_}{01}{16} & \animategraphics[autoplay,loop,width=0.2\columnwidth]{5}{images/ours-single/seq_08_recon_}{01}{16}\\
        & 26.29, 0.856 & 31.31, 0.937 & \bf 31.66, 0.940\\ 
        \includegraphics[width=0.2\columnwidth]{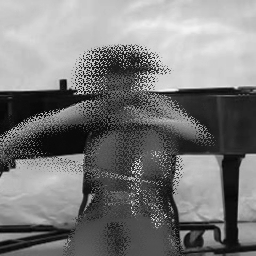}& \animategraphics[autoplay,loop,width=0.2\columnwidth]{5}{images/ours-single-interm/seq_07_interm_}{01}{16} & \animategraphics[autoplay,loop,width=0.2\columnwidth]{5}{images/ours-single-interm/seq_07_recon_}{01}{16} & \animategraphics[autoplay,loop,width=0.2\columnwidth]{5}{images/ours-single/seq_07_recon_}{01}{16}\\
        & 25.47, 0.871 & 31.02, 0.952 & \bf 31.23, 0.954\\
        \hline
    \end{tabular}
    \caption{Ablation study on architecture. The figure shows the intermediate and final reconstructions for test videos from the \textit{DNN set~\cite{yoshida2018joint}}. The final reconstructed video from the second architecture is better than the final reconstructed video from the first architecture. (videos can be viewed in Adobe Reader)}
    \label{fig:ablation}
    \end{center}
    \vspace{-22pt}
\end{figure}

\textbf{Ablation study on proposed architecture:} We explain some of the architectural choices that we made in developing our proposed network. 
We experimented with two different architectures for pixel-wise coded exposure - U-Net only, SVC(16) + U-Net, and SVC(64) + U-Net.
SVC denotes the shift-variant convolution layer~\cite{okawara2020action}, and the following value in bracket specifies the number of output channels of the SVC layer. 
In U-Net only framework, we input the coded image directly to the standard U-Net architecture, which learns the mapping to the full resolution video sequence.
In SVC(16)+U-Net, we implemented the SVC layer to produce an intermediate reconstruction from the input, followed by U-Net~\cite{ronneberger2015u} to refine the intermediate reconstruction and produce the final high-quality video.
While training the network, we supervise both the intermediate and final reconstructions using ground truth with a $0.5$ weightage for intermediate reconstruction.
In SVC(64)+U-Net, we modified the number of output channels of the SVC layer from 16 to 64. 
Therefore, instead of producing an intermediate reconstruction, the SVC layer extracts the features required to reconstruct the video. 
Here, we supervise the final reconstruction using ground truth while training.
From Table~\ref{table:ablation}, we observe that using SVC(64)+U-Net gives the best reconstruction results.
It can also be observed that using an SVC layer instead of a standard convolutional layer provides a significant improvement in performance.
The SVC layer also does not add significantly to the computational overhead. 
While, SVC(64)+U-Net model takes $0.011$s, Unet-only model takes $0.009$s per forward pass on a GPU for a $256\times 256\times 16$ video sequence.
Therefore, we choose SVC(64)+U-Net architecture as our proposed method.

\noindent\textbf{Ablation Study on C2B Input:} The advantage of using C2B exposure is that it captures the complementary information otherwise lost in pixel-wise coded exposure. 
C2B captures two coded exposure images: coded image and complement-coded image. We can obtain a fully-exposed or blurred image by adding the coded and complementary coded images.
There are two ways of representing the C2B input: a coded-complement image pair or coded-blurred image pair. 
We evaluated both the cases and determined that video reconstruction from a coded-blurred image pair performs marginally better than reconstruction from a coded-complement pair. The results are summarized in Table~\ref{table:ablation}. 

\setlength{\tabcolsep}{3pt}
\begin{table}
    \fontsize{9}{7.2}\selectfont
    \begin{center}
    \begin{tabular}{llllll}
    \hline\noalign{\smallskip}
        \multirow{2}{*}{Exposure}& &  \multicolumn{2}{c}{DNN set~\cite{yoshida2018joint}}& \multicolumn{2}{c}{GoPro set~\cite{Nah_2017_CVPR}}\\
        \noalign{\smallskip}
        && PSNR& SSIM& PSNR & SSIM\\
        \hline
        \noalign{\smallskip}
        & U-Net only & 30.68 & 0.919 & 31.27 & 0.902 \\
        Pixel-wise& SVC(16)+U-Net& 30.89& 0.921& 31.56& 0.910\\
        coded& SVC(64)+U-Net& \bf 31.14& \bf 0.925& \bf 31.76& \bf 0.914\\
        \noalign{\smallskip}
        \hline
        \noalign{\smallskip}
        \multirow{2}{*}{C2B}& coded+complement & 32.19& 0.935& 32.31& 0.919\\
        & coded+blurred & \bf 32.23& \bf 0.935& \bf 32.34& \bf 0.920\\
        \noalign{\smallskip}
        \hline
    \end{tabular}
    \end{center}
    \vspace{-15pt}
    \caption{Ablation studies on proposed architecture and C2B input. The table lists average PSNR(dB) and SSIM of reconstructed videos from \textit{DNN set~\cite{yoshida2018joint}} and \textit{GoPro set~\cite{Nah_2017_CVPR}}.{\smallskip}}
    \label{table:ablation}
    \centering
    \begin{tabular}{l|cc|cc}
         & \multicolumn{2}{c}{Noiseless} & \multicolumn{2}{|c}{Noisy($\sigma=0.01$)} \\
        \cline{2-5} 
        \noalign{\smallskip}
        Model & PSNR & SSIM & PSNR & SSIM \\
        \hline \noalign{\smallskip}
        FS (fixed) & 21.61  & 0.752 & 21.28 & 0.707 \\
        FS (optimized) & 21.72 & 0.756  & 21.42 & 0.722\\
        \hline \noalign{\smallskip}
        Pixel-wise(fixed) & 31.76 & 0.914 & 27.58  & 0.845 \\
        Pixel-wise(optimized) & 32.13 & 0.953 & 29.58 & 0.912\\
        \hline
        \noalign{\smallskip}
        C2B(fixed) & 32.34 & 0.920 & 28.22 & 0.860 \\
        C2B(optimized) & 32.59 & 0.961  & 30.06 & 0.912\\
    \end{tabular}
    \caption{
    PSNR, SSIM comparison of reconstructed for various exposure techniques for \emph{fixed} and \emph{optimized} coded mask $\phi$. We observe better reconstruction performance for \emph{Optimized} mask for both the noisy and noiseless cases.
    }
    \label{tab:learned_mask}
    \vspace{-20pt}
\end{table}
\setlength{\tabcolsep}{1.4pt}

\subsection{Learning the mask}

Jointly learning the coded mask $\phi$ and the reconstruction algorithm has been shown to provide better reconstruction results~\cite{li2020endtoend,okawara2020action,iliadis2020deepbinarymask}.
To demonstrate this, we jointly learn the coded mask $\phi$ along with our proposed learning-based reconstruction algorithm.
We add the weights of the mask $\phi$ also as trainable parameters along with the other trainable network parameters.
As the hardware sensors can use only binary mask patterns, we restrict the mask weights to be binary.
Binarization is done via thresholding the weights before each forward pass through the network.
As thresholding is non-differentiable, we follow \cite{hubara2016binarized} and use the \emph{straight-through estimator} for computing gradients.
We use a similar training scheme and training dataset as described in Sec.~\ref{sec:exptsetup}. 
The mask $\phi$ and the network are jointly trained for $16$x reconstruction for the case of FS, pixel-wise exposure, and C2B.
The trained network is evaluated on the GoPro test set, and the results are summarized in Table~\ref{tab:learned_mask}.
We observe that for both the noiseless and the noisy cases, joint optimization of the coded mask and the reconstruction algorithm provides better performance.
The gap between the fixed and optimized code is bigger for the noisy case. 

\section{Conclusion}
We propose a unified deep learning-based framework to make a fair comparison of the video reconstruction performance of various coded exposure techniques.
We make a mathematically informed choice for our framework that leads to the use of fully convolutional architecture over a fully connected one.
Extensive experiments show that the proposed algorithm performs better than previous video reconstruction algorithms across all coded exposure techniques.
The proposed unified learning framework is used to make an extensive quantitative and qualitative evaluation of the different coded exposure techniques.
From this, we observe that C2B provides the best reconstruction performance, closely followed by the single pixel-wise coded exposure technique, while FS lags far behind.
Our further analysis of C2B shows that a significant advantage is gained over pixel-wise coded exposure only when the scenes are largely static.
However, when the majority of scene points undergo motion, C2B shows only a marginal benefit over acquiring a single pixel-wise coded exposure measurement.


\section*{Code and Supplementary Material}
The official implementation of this paper with pretrained models can be found at \url{https://github.com/asprasan/unified\_framework}
The supplementary material containing the video reconstruction results can be found \href{https://drive.google.com/file/d/1GGpgzTGXnE1XzONJOv81iK23TVIfkAvB/view?usp=sharing}{here}.

\section*{Acknowldegements}
The authors would like to thank Sreyas Mohan and Subeesh Vasu for their helpful discussions.

{\small
\balance
\bibliographystyle{ieee_fullname}
\bibliography{egbib}
}

\end{document}